\documentclass[10pt]{article}

\usepackage{graphicx}

\usepackage{wds11}

\lefthead{SAGUN ET AL.}
\righthead{MIXED PHASE FORMATION}

\begin{document}

\title{Evidence of the mixed phase formation in nucleus-nucleus collisions}

\author{V. V. Sagun, K. A. Bugaev, A. I. Ivanytskyi}

\affil{Bogolyubov Institute for Theoretical Physics, Kyiv, Ukraine.}

\author{D. R. Oliinychenko}

\affil{FIAS, Frankfurt upon Main, Germany.}

\begin{abstract}
Searchers for  various  irregularities in the behavior of thermodynamic  quantities at chemical freeze-out (CFO)  are rather important in a view of experimental studies of  quark-gluon plasma  (QGP). Using the multicomponent hadron resonance gas model (HRGM), developed in
 [\markcite{{\it Sagun}, 2014}; \markcite{{\it Bugaev et al.(1)}, 2015}], we performed  a high-quality fit of 111 hadronic ratios measured for 14 values of the center of mass collision energies between 2.7 GeV and 200 GeV with the overall fit quality $\chi^2/dof  \simeq 0.95$. 
In addition to previously reported singularities [\markcite{{\it Bugaev et al.(1)}, 2015}]
at CFO we found that the hadron yield ratios $\frac{\Lambda}{p}$, $\frac{K^{+}}{p}$, $\frac{K^{+}}{\Lambda}$, $\frac{\Omega^{-}}{p}$ and $\frac{\Xi^{-}}{p}$ measured in central nuclear collisions  demonstrate a significant change of slope in  the same   range of  center of mass collision energy    $\sqrt{s_{NN}} = 4.3-4.9$ GeV   [\markcite{{\it Bugaev et al.(2)}, 2015}].  This  change of slopes  is accompanied by  a dramatic increase of resonance decays at CFO. 
 Also  at CFO  the trace anomaly   and baryonic density demonstrate the  pronounced  peaks  at the collision energy $\sqrt{s_{NN}} = 4.9 $ GeV.  We argue that all these and previously found irregularities provide an evidence for  the QGP formation in nuclear collisions
 at about $\sqrt{s_{NN}} = 4.9 $ GeV.
\\
 
\end{abstract}

\begin{article}

\section{Introduction}

Experimental searches for the QGP  in heavy-ion collisions  cannot be completed successfully  without  reliable and  justified signals of its formation. Although some irregularities, known as the Kink  [\markcite{{\it Gazdzicki}, 1995}], the Strangeness Horn [\markcite{{\it Gazdzicki et al.}, 1999}] and the Step [\markcite{{\it Gorenstein et al.}, 2003}],  are widely  considered as  the signals of the onset of deconfinement [\markcite{{\it Gazdzicki et al.}, 2011}], their relation to the QGP-hadron mixed phase is far from being clear. 
Therefore, the development of realistic models which are  able to accurately describe the existing experimental data and to  provide us with the reliable information about the late  stages of the heavy-ion collision process is absolutely necessary. 

 The high quality description of data achieved recently for  111 independent hadron yield ratios measured at midrapidity  in central nucleus-nucleus collisions for 14 values of the center of mass energies undoubtedly proves that the  HRGM with the multicomponent  
 hard-core repulsion   is a precise and a sensitive tool  of  heavy ion collision phenomenology [\markcite{{\it Sagun}, 2014}]. In contrast to other existing  versions of HRGM [\markcite{{\it Andronic et al.}, 2006};  \markcite{{\it Bugaev et al.}, 2013}] the present one accounts for the hard-core repulsion using different hard-core radii for pions, $R_{\pi}$, kaons, $R_K$,  $\Lambda$-hyperons, $R_\Lambda$, other mesons, $R_m$, and other baryons, $R_b$. With such a model we are able to fit the experimental multiplicities measured at AGS for $\sqrt s_{NN}=2.7, 3.3, 3.8, 4.3, 4.9$ GeV, the NA49 data measured at SPS energies  $\sqrt s_{NN}=6.3, 7.6, 8.8, 12.3, 17.3$ GeV and the STAR data measured  at RHIC energies  $\sqrt s_{NN}=9.2, 62.4, 130, 200$ GeV  with the highest quality $\chi^2/dof  \simeq 0.95$ [\markcite{{\it Sagun}, 2014}; \markcite{{\it Bugaev et al.(1)}, 2015}]. Therefore, using the HRGM with the multicomponent hard-core repulsion we can study thermodynamics of strongly interacting matter at CFO   with very  high confidence. 
 
  Using the   multicomponent version of  HRGM   [\markcite{{\it Sagun}, 2014}; \markcite{{\it Bugaev et al.(1)}, 2015}] 
 we previously found  the set  of strongly correlated quasi-plateaus  in the collision energy dependence of the entropy per baryon and of the pion number (both the thermal and total) per baryon at CFO   [\markcite{{\it Bugaev et al.(1)}, 2015}].  Such strongly correlated plateaus were predicted a long time ago as a signal of  QGP-hadron mixed phase formation  [\markcite{{\it Bugaev et al.}, 1989}; \markcite{{\it Bugaev et al.}, 1990}; \markcite{{\it Bugaev et al.}, 1991}].  Hence, it was argued   that the observed  dramatic changes in the system properties seen in the narrow collision energy range $\sqrt{s_{NN}} = 4.3-4.9 $ GeV  evidence for  the QGP formation in nuclear collisions  [\markcite{{\it Bugaev et al.(1)}, 2015}].
 
With the help of  this  HRGM we also found the new irregularities related to deconfinement.  The most spectacular of them  is a sudden jump of the pressure $p$ at CFO in the narrow range of center of mass  collision energies $\sqrt{s_{NN}} = 4.3-4.9$ GeV [\markcite{{\it Bugaev et al.(3)}, 2015}, \markcite{{\it Bugaev et al.(4)}, 2015}]. The observed pressure jump at CFO is so strong, that the effective number of degrees of freedom, $p/T^4$, increases by 70\%, while the collision energy in the center of mass system changes by 15\% and   while the CFO temperature $T$  changes by 30\%. Below we discuss 
other  irregularities and possible  signals of the mixed phase formation in nuclear collisions which  are of great importance  for the success  of the planned heavy-ion collision  experiments at JINR-NICA  and  GSI-FAIR. 

\section{HRGM with multicomponent hard-core repulsion}

Let us consider  the Boltzmann gas of $N$ hadron species in a volume $V$ that has  the temperature $T$, the baryonic chemical potential $\mu_B$, the  strange chemical potential $\mu_S$ and the chemical potential of the isospin third component $\mu_{I3}$. Such an approach  is based on the assumption of  local thermal and chemical equilibrium at CFO. Hence the hadron yields  produced in the collisions of large atomic nuclei can be found using the grand canonical valuables. The system  pressure $p$ and the $K$-th charge density $n^K_i$ ($K\in\{B,S, I3\}$) of the  i-th hadron sort are given by the expressions  
\begin{eqnarray}\label{EqI}
p = \sum_{i=1}^N p_i \,,   \quad \quad 
n_i^K = \frac{Q_i^{K} p_i}{T+ \frac{\sum_{jl} p_j b_{jl} p_l }{p}}, 
\end{eqnarray}
where $b_{ij} = \frac{2\pi}{3}(R_i+R_j)^3$ is a symmetric  matrix of the second  virial coefficients, {and $R_j$ is the hard-core radius of hadron of sort $j$.
}
The equation of state of the system is written in terms of partial pressures $p_i$
\begin{eqnarray}\label{EqII}
p_i = T \phi_i (T)\,   \exp \left[ \frac{\mu_i - 2 \sum_{j} p_j b_{ji} + \sum_{jl} p_j b_{jl}p_l/p }{T} \right]  \,, \quad \quad \hspace*{-4mm}\phi_i (T)  = \frac{g_i}{(2\pi)^3}\int \exp\left[-\frac{\sqrt{k^2+m_i^2}}{T} \right]d^3k \,.~
\end{eqnarray}
Here   the full chemical potential of the $i$-th hadron sort $\mu_i \equiv Q_i^B \mu_B + Q_i^S \mu_S + Q_i^{I3} \mu_{I3}$ is expressed in terms of the corresponding charges $Q_i^K$  and their  chemical potentials,  $ \phi_i (T) $ denotes 
the thermal particle  density of  the $i$-th hadron sort of mass $m_i$ and degeneracy $g_i$, and  $\xi^T$  denotes  the row of  variables $\xi_i$.  For each collision energy the fitting parameters are the temperature $T$, the baryonic chemical potential $\mu_B$ and the chemical potential of the third projection of  isospin  $\mu_{I3}$, whereas the strange chemical potential  $\mu_S$ is found from the   condition of  vanishing strangeness.

In order to  account for the possible  strangeness non-equilibrium  we employ  the  $\gamma_s$ factor  [\markcite{{\it Rafelski et al.}, 1982}]  by replacing $\phi_i$ in Eqs. (\ref{EqII})  as $\phi_i(T) \to \phi_i(T) \gamma_s^{s_i}$. Here $s_i$ is a number of strange valence quarks plus number of strange  valence anti-quarks.
The 
width correction is taken into account by averaging the Boltzmann exponent  with the  Breit-Wigner distribution.  As a result, the modified thermal particle density of $i$-th hadron  sort  acquires  the form
\begin{eqnarray}
\label{EqV}
\int \exp\left(-\frac{\sqrt{k^2+m_i^2}}{T} \right)d^3k \rightarrow
 \frac{\int^{\infty}_{M_{0}} \frac{dx}{(x-m_{i})^{2}+\Gamma^{2}_{i}/4} \int \exp\left(-\frac{\sqrt{k^2+x^2}}{T} \right)d^3k }{\int^{\infty}_{M_{0}} \frac{dx}{(x-m_{i})^{2}+\Gamma^{2}_{i}/4}} \,. 
\end{eqnarray}
Here $m_i$ denotes the  mean mass of  hadron and $M_0$ stands for the threshold in the dominant decay channel. The main advantages of this approximation  are a simplicity of  its realization and a clear way to account for the finite  width of hadrons. 
 
The effect of resonance decay $Y \to X$ on the final hadronic multiplicity is taken into account as $n^{fin}(X) = \sum_Y BR(Y \to X) n^{th}(Y)$, where $BR(X \to X)$ = 1 for the sake of convenience. 
The masses, the  widths and the strong decay branchings $BR(Y \to X)$ of all experimentally known hadrons  were  taken from the particle tables  used  by  the  thermodynamic code THERMUS [\markcite{{\it Wheaton et al.}, 2009}]. The detailed description of the analyzed   data sets  together with the fit procedure can be found in  [\markcite{{\it Sagun}, 2014}].
\begin{table}[!h]
{\footnotesize
\begin{center}
\hspace*{0.3cm}
\begin{tabular}{|c|c|c|c|c|c|c|c|c|c|c|c|c|c|c|}
\hline
$\sqrt{s_{NN}}$ (GeV) & 2.7  & 3.3  & 3.8  & 4.3  & 4.9 & 6.3 & 7.6 & 8.8 & 9.2 & 12   & 17.0 & 62.4 & 130.0 &200.0 \\
\hline
$\chi^2$              & \hspace*{-0.24cm} 0.428 \hspace*{-0.24cm}& \hspace*{-0.24cm} 0.052 \hspace*{-0.24cm}& \hspace*{-0.24cm} 0.048 \hspace*{-0.24cm} & \hspace*{-0.24cm} 0.306 \hspace*{-0.24cm} & \hspace*{-0.24cm} 0.149 \hspace*{-0.24cm} & \hspace*{-0.24cm} 5.936 \hspace*{-0.24cm} & \hspace*{-0.24cm} 4.201 \hspace*{-0.24cm} & \hspace*{-0.24cm} 3.735 \hspace*{-0.24cm} & \hspace*{-0.24cm} 0.002 \hspace*{-0.24cm} & \hspace*{-0.24cm} 11.992 \hspace*{-0.24cm} & \hspace*{-0.24cm} 13.633 \hspace*{-0.24cm} & \hspace*{-0.24cm} 0.2807 \hspace*{-0.24cm} & \hspace*{-0.24cm} 4.337 \hspace*{-0.24cm} &\hspace*{-0.24cm} 6.898 \hspace*{-0.24cm}\\
\hline
\hspace*{-0.24cm} ratios number \hspace*{-0.24cm} & 4    & 5    & 5    & 5    & 8   & 9   &  10 & 11  &  5  & 10   &  13  &  5   &  11 & 10\\  
\hline
\end{tabular}
\end{center}
} 
\caption{\label{tab:bolts} The quality of data description achieved  for the center of mass collision energy $\sqrt{s_{NN}}$.}
\end{table}

 The best  global fit  of all hadronic multiplicities corresponds to  $R_b$ = 0.355 fm, $R_m$ = 0.4 fm, $R_{\pi}$ = 0.1 fm,  $R_K$ = 0.38 fm and $R_\Lambda = 0.11$ fm with the quality  $\chi^2/dof \simeq 0.95$ [\markcite{{\it Sagun}, 2014}]. 
 These hard-core radii  were chosen, since the most abundant particle which are measured for  all collision energies are pions, kaons, protons and (anti)$\Lambda$ hyperons.  As a result,  this set of hard-core radii allowed us for the first time to 
 simultaneously describe the peaks in $K^+/\pi^+$ and $\Lambda/\pi^-$ ratios [\markcite{{\it Sagun}, 2014}, \markcite{{\it Bugaev et al.(1)}, 2015}] without spoiling the  high quality description of all other ratios. 
  Note that for more than a decade these particle yield  ratios, known  as  the Strangeness Horn  $K^+/\pi^+$ and the Lambda hyperon horn $\Lambda/\pi^-$,  
 were  the most problematic ones for the traditional  thermal models  [\markcite{{\it Andronic et al.}, 2006}].
In Figure~1 we compare  these experimental  ratios and their fits  achieved by the multicomponent HRGM.
The obtained  quality of  fit  of   independent particle ratios  
 for each  center of mass collision energy point     is presented in Table  1.

\subsection{Irregularities at chemical freeze-out}

\begin{figure}[!]
\begin{center}
\includegraphics[width=38.1pc]{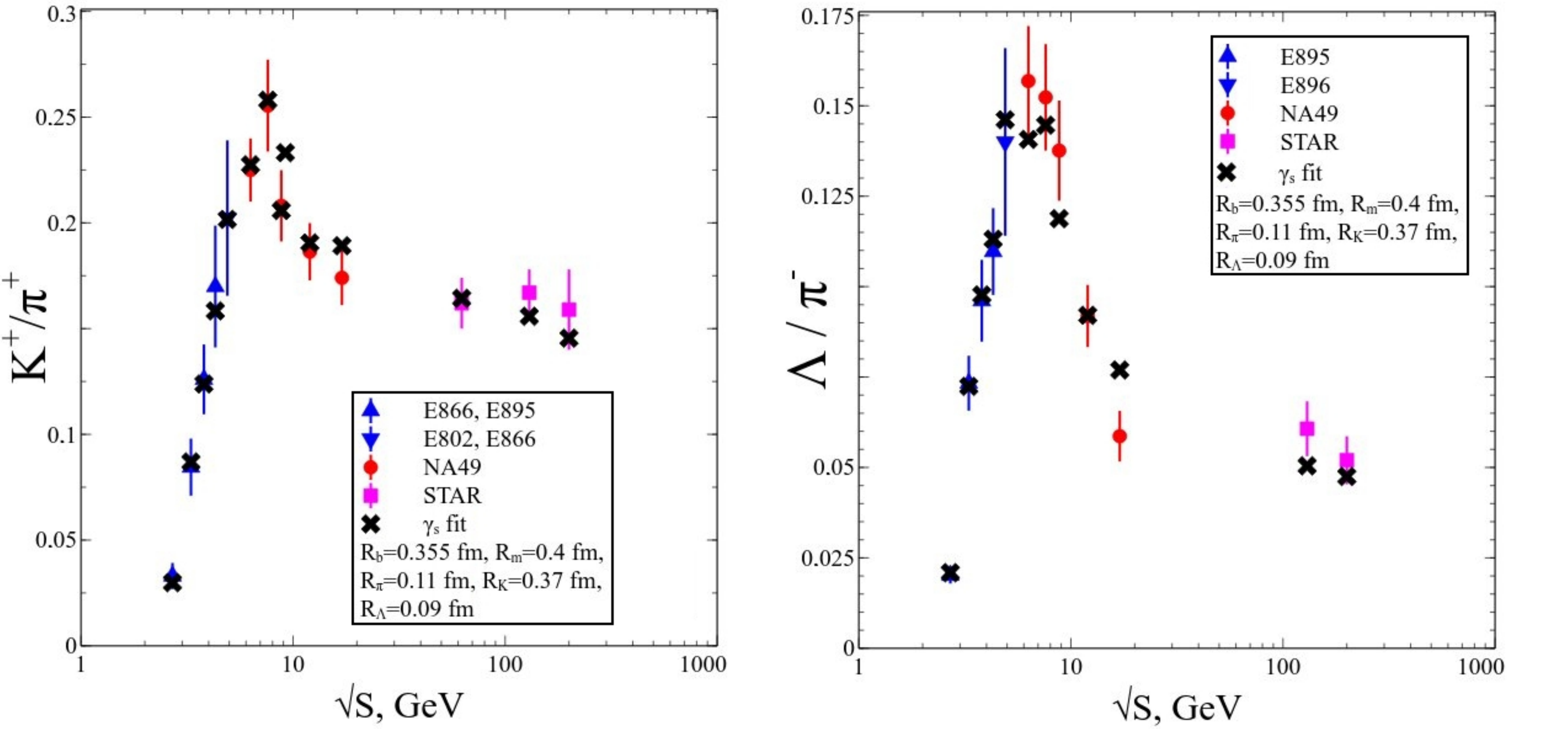}
\end{center}
\caption{Collision energy dependence of $\frac{K^+}{\pi^+}$  (left panel, $\chi^2/dof = 3.9/14$) and $\frac{\Lambda}{\pi^{-}}$  (right panel, $\chi^2/dof = 10.2/12$) hadron yield ratios.}
\end{figure}

It is necessary to stress that with the help of the present formulation of
 HRGM  a  plateau in the collision energy dependence of  the thermal pion number per baryon and the quasi-plateaus  in the entropy per baryon   and in the total pion number per baryon  were found  recently in [\markcite{{\it Bugaev et al.(1)}, 2015}].  
A simultaneous  appearance of these quasi-plateaus   was predicted a long time ago in  [\markcite{{\it Bugaev et al.}, 1989}; \markcite{{\it Bugaev et al.}, 1990}; \markcite{{\it Bugaev et al.}, 1991}] to be  a signal of the QGP-hadron mixed phase formation.
Their  appearance is a manifestation of anomalous thermodynamic properties of the mixed phase. 
 
The validity of  this signal   is strongly supported by an existence of the sharp peak of the trace anomaly at  $\sqrt{s_{NN}} = 4.9 $ GeV (see the right panel of Figure~2). 
Note that the inflection point/maximum of the trace anomaly is traditionally used in lattice QCD to determine the pseudocritical temperature of the cross-over transition \markcite{{\it Borsanyi et al.}, 2012}].
Furthermore,
our simulations of the generalized shock adiabat show that this  peak of $\delta$ at CFO is a consequence of  the peak of 
trace anomaly on the shock adiabat  located exactly at the boundary of the mixed phase and QGP.
From the expression for trace anomaly

\begin{figure}[!h]
\begin{center}
\includegraphics[width=38.1pc]{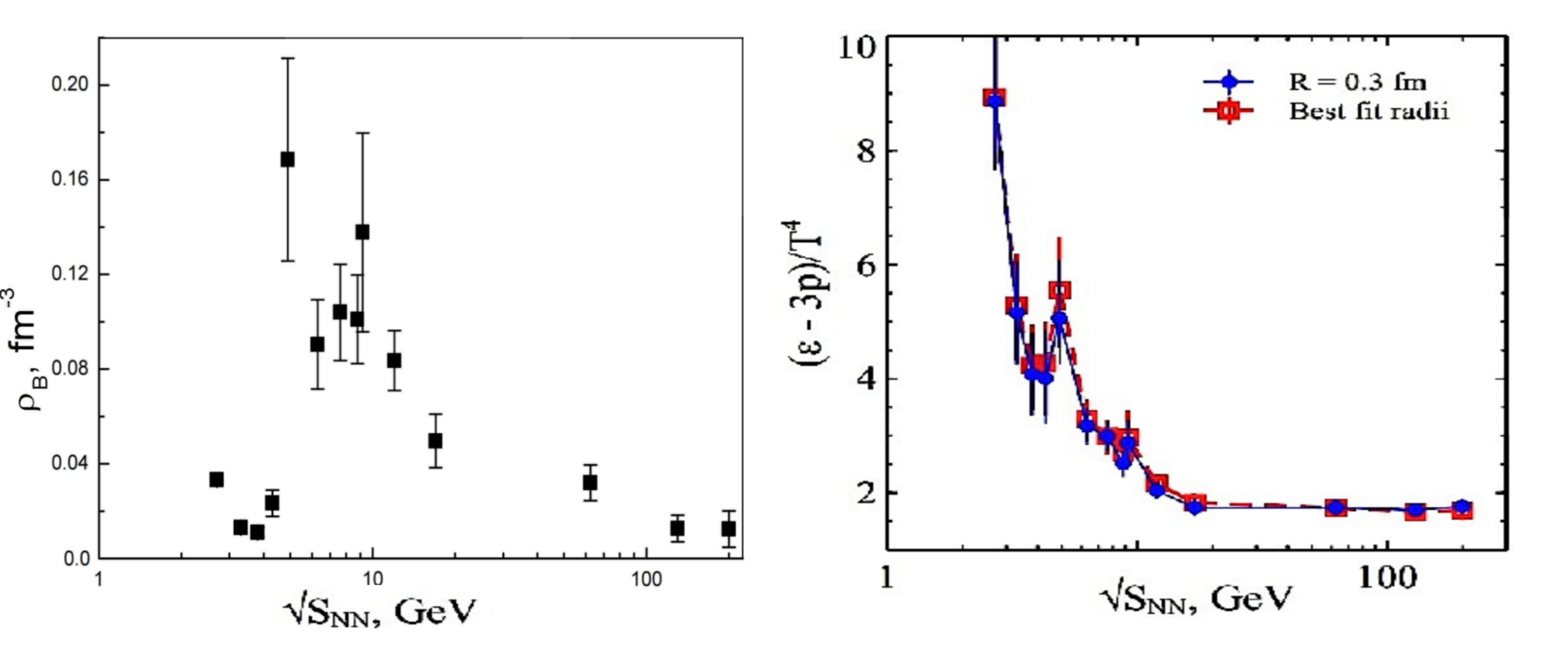}
\end{center}
\caption{Baryonic charge density (left panel) and trace anomaly (right panel) as functions of collision energy at CFO.}
\end{figure}

\begin{eqnarray}
\delta=\frac{\varepsilon - 3p}{T^4} \simeq \frac{s}{T^3} \left(1+ \frac{\mu_{B}}{T} \frac{\rho_{B}}{s} \right) - 4 \frac{p}{T^4}, 
\end{eqnarray}
one can show [\markcite{{\it Bugaev et al.(4)}, 2015}] that 
the  strong increase of $\delta$ when  the collision energy changes from $\sqrt{s_{NN}} = 4.3$ GeV to $\sqrt{s_{NN}} = 4.9 $ GeV is provided by a strong jump of the effective number of degrees of freedom $\frac{s}{T^3} $ on this interval. Here $\varepsilon$ is the energy density and $s$ is the entropy density. 

Also  at $\sqrt{s_{NN}} = 4.9 $ GeV we observe   a  sharp peak of 
the baryonic density at CFO (see the left panel of Figure~2), which is a consequence of the baryonic density peak at the generalized shock adiabat existing at the boundary between the mixed phase and QGP.

\begin{figure}[!h]
\begin{center}
\includegraphics[width=22.1pc]{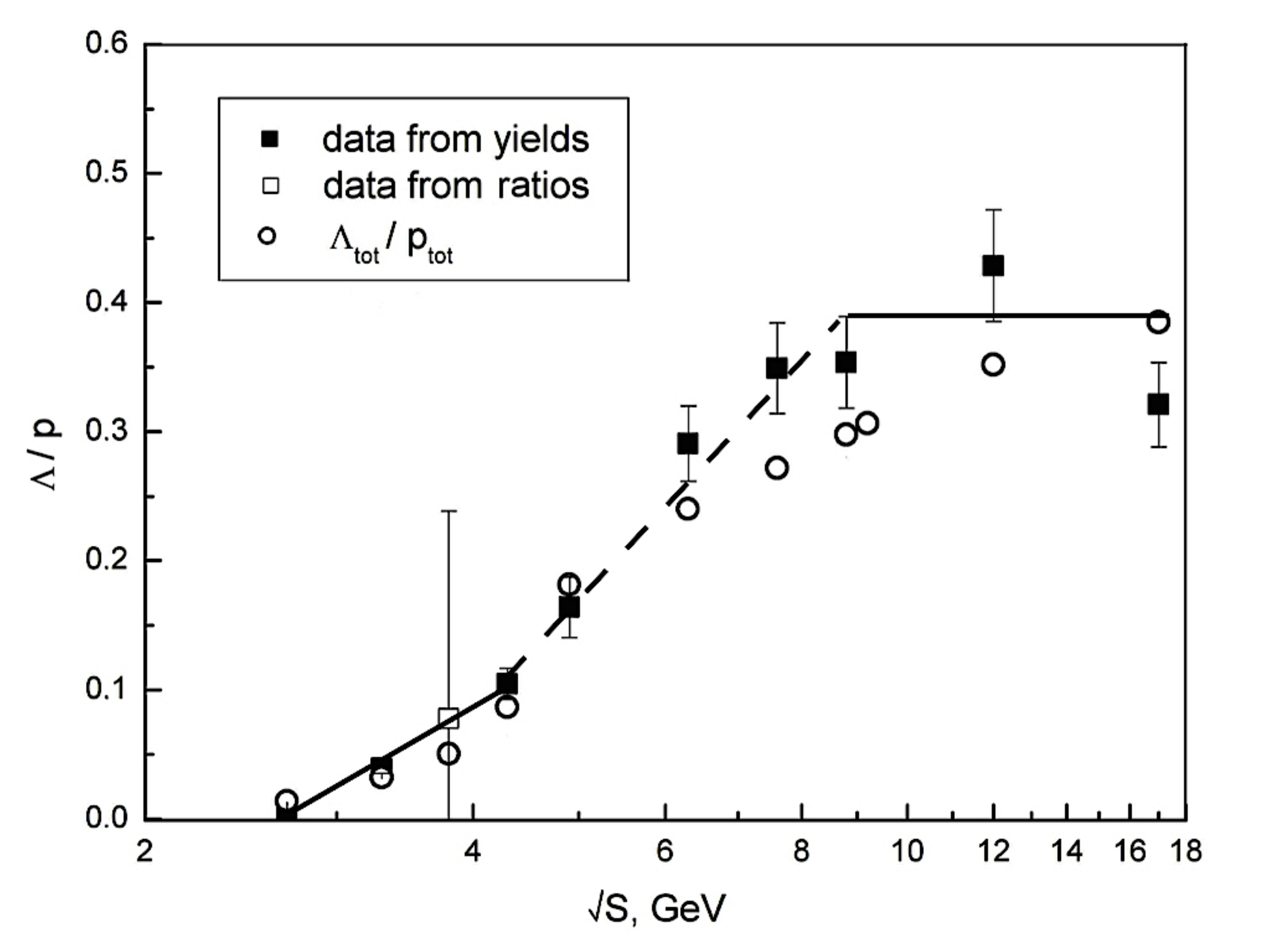}
\end{center}
\caption{The center of mass collision energy dependence of the $\frac{\Lambda}{p}$ ratio obtained within the present HRGM. The lines are given to guide the eye.}
\end{figure}

\begin{figure}[!h]
\begin{center}
\includegraphics[width=38.1pc]{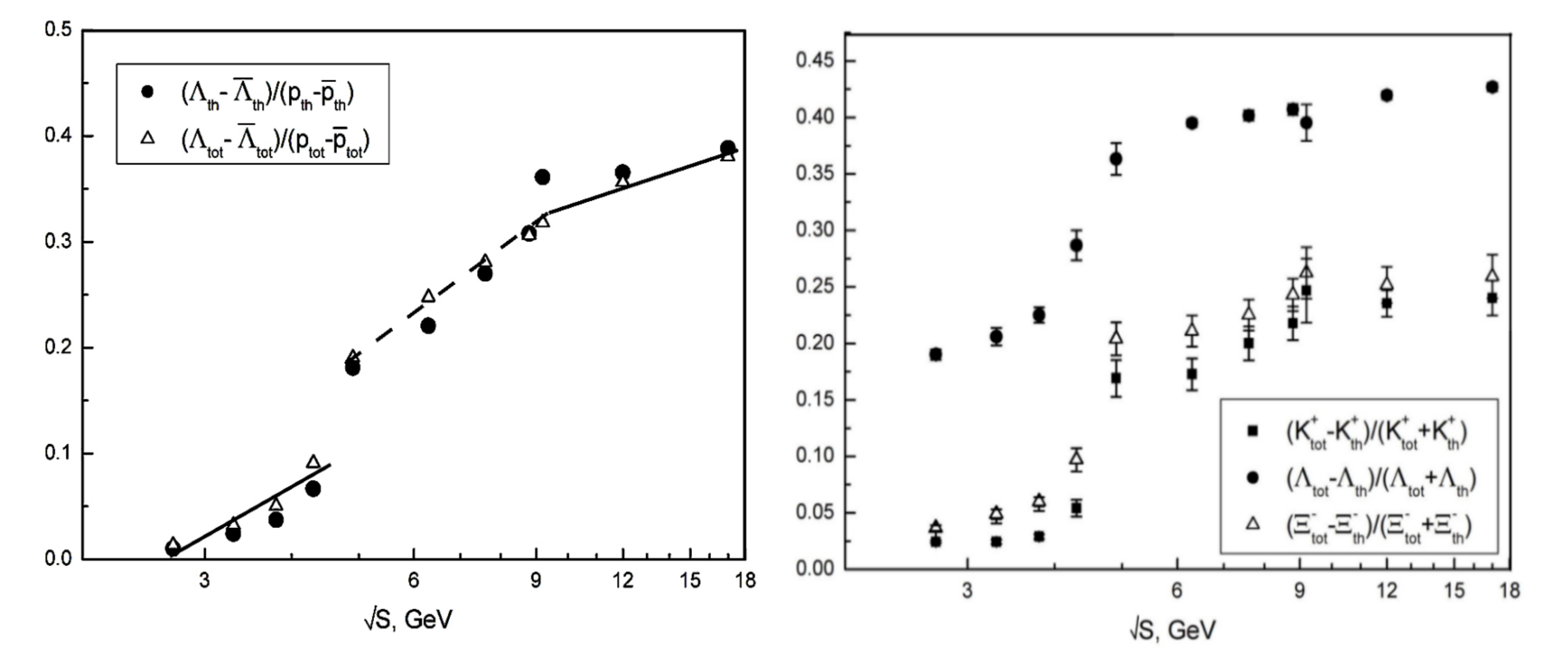}
\end{center}
\caption{Center of mass collision energy $\sqrt{s_{NN}}$ dependence of $\frac{\Delta \Lambda}{\Delta p}$ (left panel) and the asymmetry  between total and thermal multiplicities for kaons (squares), $\Lambda$  (circles) and and $\Xi^{-}$ (triangles) hyperons (right panel) obtained within the present HRGM. The lines are given to guide the eye.
}
\end{figure}

The other evidence of a possible deconfinement transition  between the collision  energies $\sqrt{s_{NN}} = 4.3 $ and $4.9 $ GeV is provided by  the  sudden increase of $\frac{\Lambda}{p}$ slope at $\sqrt{s_{NN}} = 4.3 $ GeV (see Figure~3). The same behavior was found for $\frac{K^{+}}{p}$, $\frac{\Omega^{-}}{p}$ and $\frac{\Xi^{-}}{p}$ ratios. This behavior  can be naturally explained [\markcite{{\it Bugaev et al.(4)}, 2015}] by the idea suggested in [\markcite{{\it Rafelski et al.}, 1982}]  that the mixed phase formation can be detected  by a rapid increase in the number of strange quarks per number of  light quarks. Evidently, the   $\Lambda/p$    ratio is a  convenient indicator because at low collision energies the $\Lambda$-hyperons are generated in collisions of nucleons.  Moreover, such a ratio does not depend on baryonic chemical potential, since both the protons and $\Lambda$-hyperons have the same baryonic charge. 
As it is seen from the  Figure~3, this mechanism works up to  $\sqrt{s_{NN}}=4.3$ GeV,
while an appearance of the mixed phase  leads to an  increase of  the number of strange quarks and antiquarks  due to the annihilation of light quark-antiquark and gluon pairs.

A similar behavior with a clear jump at the energies $\sqrt{s_{NN}}=4.3-4.9$ GeV we predict for  the collision energy dependence  of  the ratio  $\frac{\Delta \Lambda}{\Delta\, p} \equiv \frac{\Lambda - \bar\Lambda}{p - \bar p}$  which is  shown in the left panel of Figure~4. Since the observed jump of this ratio  is located in  the collision energy region of the mixed phase formation (i.e. with a first order phase transition), then a change of  its   slope at $\sqrt{s_{NN}} = 9.2$ GeV  can be naturally associated with a weak first order or a second order phase transition [\markcite{{\it Bugaev et al.(4)}, 2015}]. 
Note that this hypothesis is well supported by the second peak of the trace anomaly existing at  $\sqrt{s_{NN}} = 9.2$ GeV (see the right panel of Figure~2) and by the 
second set of quasi-plateaus found for  $\sqrt{s_{NN}} \in [7.6; 9.2]$ GeV [\markcite{{\it Bugaev et al.(1)}, 2015}]. 

The observed irregularities are also accompanied by a sudden increase of the strange  particle decays at CFO which is also seen in  the same collision energy range $\sqrt{s_{NN}} \in [4.3; 4.9]$ GeV.  The  asymmetries   between  the total and the  thermal particle yields $\frac{K^{+}_{tot}-K^{+}_{th}}{K^{+}_{tot}+K^{+}_{th}}$, $\frac{\Lambda_{tot}-\Lambda_{th}}{\Lambda_{tot}+\Lambda_{th}}$, $\frac{\Xi^{-}_{tot}-\Xi^{-}_{th}}{\Xi^{-}_{tot}+\Xi^{-}_{th}}$ are shown in  the right panel of Figure~4.

\section{Conclusions}

 In addition to the irregularities discussed previously in [\markcite{{\it Bugaev et al.(2)}, 2015}; \markcite{{\it Bugaev et al.(3)}, 2015}] here we present the new set of quantities which demonstrate a significant change in the same narrow range of the center of mass collision energy $\sqrt{s_{NN}}=4.3-4.9$ GeV. Using the multicomponent HRGM we achieved  a nearly perfect  description of the $\frac{\Lambda}{p}$ ratio for the collision energies $\sqrt{s_{NN}}$ = 2.7-7.6 GeV.  This result gives us a high confidence in the prediction  that the ratios $\frac{\Lambda- \bar{\Lambda}}{p-\bar{p}}$ and $\frac{K^{+}- \bar{K^{+}}}{p-\bar{p}}$ 
 can serve as the reliable indicators of deconfining phase transformation in heavy ion collisions. These irregularities are also accompanied by the jump of asymmetry between the  total  and  the thermal particle yields $\frac{K^{+}_{tot}-K^{+}_{th}}{K^{+}_{tot}+K^{+}_{th}}$, $\frac{\Lambda_{tot}-\Lambda_{th}}{\Lambda_{tot}+\Lambda_{th}}$, $\frac{\Xi^{-}_{tot}-\Xi^{-}_{th}}{\Xi^{-}_{tot}+\Xi^{-}_{th}}$, indicating a significant role of the strange  particle decays at CFO. 
 In addition, we found the strong sharp peaks in the trace anomaly $\delta=\frac{\varepsilon - 3p}{T^4} $ and  in the baryonic charge density at  $\sqrt{s_{NN}} = 4.9 $ GeV, 
 which are related to the peaks of corresponding quantities on the generalized shock adiabat located  at the boundary between the mixed phase and QGP. 
 
 Therefore, we conclude that a dramatic change of  the system properties seen in the narrow collision energy range $\sqrt{s_{NN}} = 4.3-4.9 $ GeV opens entirely new possibilities for the  FAIR and NICA experiments. We guess  that  the second set of peaks of these quantities which  we observe at the collision energy  $\sqrt{s_{NN}} = 9.2 $ GeV may evidence for another phase transformation in heavy ion collisions, but to get more definite conclusions about them  we need more experimental data measured with essentially higher precision.  We hope that the future experiments at NICA and FAIR will provide us with such data.

\acknowledgments 
The authors thank D. B. Blaschke, I. N. Mishustin, D. H. Rischke  and  L. M. Satarov for the fruitful discussions. The  present  work  was  supported in part by the National Academy of Sciences of Ukraine.

\end{article}

\end{document}